\documentclass[12pt,thmsa]{article}
\usepackage{amsfonts}

\usepackage{graphicx}
\usepackage{sw20aip}

\input{tcilatex}
\begin{document}

\author{Manfred Bucher \and Physics Department, California State University, Fresno
\and Fresno, CA 93740-8031, USA}
\title{Bohr model without quantum jumps }
\date{November 17, 2005}
\maketitle

\begin{abstract}
Omission of Bohr's second postulate permits a derivation of spectral
intensity with transition amplitudes $X_{nn^{\prime }}=r_{B}\left( n^{\prime
3}-n^{3}\right) /3.$ The transition amplitudes serve as upper bounds to
quantum mechanical matrix elements. They also provide insight into the
latter in terms of Sommerfeld ellipses and transition trajectories. The
speed of a nascent photon in the region of the electron transition is
addressed and the orbit concept is reinterpreted.

\bigskip \bigskip \noindent PACS numbers: 31.10.+z, 32.30.-r, 32.70.-n,
03.65.Sq
\end{abstract}

\section{INTRODUCTION}

The Bohr model of the hydrogen atom can be regarded as the greatest coup in
quantum physics. With bold assumptions it derives in a few, strikingly
simple steps a frequency formula that historically breached the
long-standing mystery of the spectral lines and provided a key to the
structure of the atom. The Bohr model is not without shortcomings though.
Chief among them is its silence on the brightness of spectral lines. The
shortcomings have led to the demise of the Bohr model and its elaboration by
Sommerfeld---the ``old quantum theory''---and the subsequent rise of quantum
mechanics.

Despite its limitations the Bohr model is still taught in introductory
physics for historical and conceptual reasons and the simple derivation of
energy levels and radiation frequencies. The model is based on two
postulates---stationary states\cite{1} and quantum leaps---and the specific
assumption of circular electron orbits. The stationary states are fixed with
quantization conditions\cite{2}, leading to orbit size,

\begin{equation}
r_{n}=r_{{\small B}}n^{2},  \tag{1}
\end{equation}

\noindent orbit energy,

\begin{equation}
E_{n}=-\frac{R_{y}}{n^{2}},  \tag{2}
\end{equation}

\noindent and orbital frequency,

\begin{equation}
f_{n}=\frac{2R_{y}}{h}\frac{1}{n^{3}},  \tag{3}
\end{equation}

\noindent all dependent on the quantum number $n=1,2,3,...$. Here $
r_{B}=h^{2}/4\pi ^{2}me^{2}$ is the Bohr radius, $R_{y}=2\pi
^{2}me^{4}/h^{2} $ the Rydberg energy, $e$ the elementary charge, $m$ the
electron mass, and $h$ is Planck's constant.

The Bohr model treats the transition of the electron from orbit $n$ to $%
n^{\prime }$ as a ``quantum leap'' with the difference in orbit energy
accounting for the energy $\epsilon $\ of an emitted or absorbed photon. The
Planck-Einstein relation associates $\epsilon $\ with the radiation
frequency $f_{nn^{\prime }}$,

\begin{equation}
E_{n^{\prime }}-E_{n}=\epsilon =hf_{nn^{\prime }}.  \tag{4}
\end{equation}

\noindent Combining Eqs. (4) and (2) gives the Balmer formula,

\begin{equation}
f_{nn^{\prime }}=\frac{R_{y}}{h}(\frac{1}{n^{2}}-\frac{1}{n^{\prime 2}}), 
\tag{5}
\end{equation}

\noindent in terms of quantum numbers and fundamental constants.\cite{3}

The connection of the ``quantum-leap world'' inside the atom with classical
electrodynamics outside is established at the ``rim'' of the atom, that is,
for very large quantum numbers compared to the transition, $n$ and $%
n^{\prime }\gg \Delta n=n^{\prime }-n$. In this limit Eq. (5) can be
approximated,

\begin{equation}
f_{nn^{\prime }}=\frac{R_{y}}{h}\frac{n^{\prime 2}-n^{2}}{n^{2}n^{\prime 2}}=%
\frac{R_{y}}{h}\frac{(n^{\prime }+n)(n^{\prime }-n)}{n^{2}n^{\prime 2}}%
\approx \frac{R_{y}}{h}\frac{2n\Delta n}{n^{4}}=f_{n}\Delta n.  \tag{6}
\end{equation}

\noindent Here Eq. (3) has been used to invoke the orbital frequency $f_{n}$
. In a transition between high neighbor orbits, $\Delta n=1$, the radiation
frequency due to the quantum jump of the electron becomes practically equal
to the electron's orbital frequency, $f_{nn^{\prime }}\cong f_{n}$. The
limiting procedure in Eq. (6), whereby the quantum realm and the macroscopic
regime merge, is called Bohr's \textit{correspondence principle}.

\section{NO QUANTUM LEAPS}

What happens if we keep Bohr's first postulate---the (quantized) stationary
states---but drop the second postulate---the quantum leaps? We then assume
that a transition from quantum state $n$ to $n^{\prime }$ is a process of 
\textit{continuously changing action}, denoted by a continuous quantum
variable $\tilde{n}$. In this view the transition of the electron from orbit 
$n$ to $n^{\prime }$ is an intermediate process with intermittent orbital
frequency $f(\tilde{n})=(2R_{y}/h)/\tilde{n}^{3}$ and intermittent radius $r(%
\tilde{n})=r_{B}\tilde{n}^{2}$ between $r_{n}$ and $r_{n^{\prime }}$. The
frequency associated with the quantum transition is obtained by summation of
infinitesimal changes of the orbital frequency,

\begin{equation}
f_{nn^{\prime }}=\int_{n}^{n^{\prime }}f(\tilde{n})d\tilde{n}=\frac{2R_{y}}{h%
}\int_{n}^{n^{\prime }}\tilde{n}^{-3}d\tilde{n}=-\left. \frac{1}{2}\frac{%
2R_{y}}{h}\tilde{n}^{-2}\right| _{n}^{n^{\prime }}=\frac{R_{y}}{h}(\frac{1}{
n^{2}}-\frac{1}{n^{\prime 2}}).  \tag{5$^{\prime }$}
\end{equation}

\noindent The result is the Balmer formula, as in Eq. (5)\cite{4}.

For very large quantum numbers, $n$ and $n^{\prime }\gg \Delta n$, the
orbital frequency $f_{n}$ changes relatively little with increasing $n$ so
that the integral of the transition frequency, Eq. (5'), between quantum
states $n$ and $n^{\prime }=n+\Delta n$ can be approximated,

\begin{equation}
f_{nn^{\prime }}=\int_{n}^{n^{\prime }}f(\tilde{n})d\tilde{n}\approx
f_{n}\int_{n}^{n^{\prime }}d\tilde{n}=f_{n}\Delta n.  \tag{6$^{\prime }$}
\end{equation}

\noindent The result is the correspondence principle, as in Eq. (6).

The projection of the electron motion in orbit $n$ onto an axis through the
nucleus can be considered an oscillating dipole,

\begin{equation}
p_{n}(t)=-eX_{n}cos(2\pi f_{n}t),  \tag{7}
\end{equation}

\noindent with amplitude $X_{n}=r_{n}$ from Eqs. (1) and frequency $f_{n}$
from Eq. (3). By classical electrodynamics\cite{5}$^{,}$\cite{6} the
instantaneous radiative power of an oscillating dipole is

\begin{equation}
S(t)=\frac{2\mathbf{\ddot{p}}^{2}}{3c^{3}},  \tag{8}
\end{equation}

\noindent where the double dot indicates the second derivative. No such
radiation,

\begin{equation}
S_{n}(t)=\frac{2}{3c^{3}}\left( \frac{d^{2}p_{n}}{dt^{2}}\right) ^{2}=\frac{
32\pi ^{4}e^{2}}{3c^{3}}f_{n}^{4}X_{n}^{2}cos^{2}(2\pi f_{n}t),  \tag{9}
\end{equation}

\noindent occurs for the dipole $p_{n}(t)$, Eq. (7), due to its postulated
stationary-state motion.\cite{7} However, radiation of frequency $
f_{nn^{\prime }}$, Eq. (5$^{\prime }$), is emitted or absorbed when a
quantum \textit{transition} occurs. To this end we replace in Eq. (9) the
orbital quantum number $n$ by the quantum-number pair $nn^{\prime }$ for the
transition. Taking the time average over a period, $\left\langle
cos^{2}...\right\rangle =1/2$, the \textit{average} radiative power becomes

\begin{equation}
\left\langle S_{nn^{\prime }}\right\rangle =\frac{16\pi ^{4}e^{2}}{3c^{3}}%
f_{nn^{\prime }}^{4}X_{nn^{\prime }}^{2}.  \tag{10}
\end{equation}

We determine the transition amplitude $X_{nn^{\prime }}$, in analogy to the
transition frequency $f_{nn^{\prime }}$, via infinitesimal increments of the
intermittent radius $r(\tilde{n})$,

\begin{equation}
X_{nn^{\prime }}=\int_{n}^{n^{\prime }}r(\tilde{n})d\tilde{n}=r_{%
{\scriptsize B}}\int_{n}^{n^{\prime }}\tilde{n}^{2}d\tilde{n}=\frac{1}{3}r_{%
{\scriptsize B}}(n^{\prime 3}-n^{3}).  \tag{11}
\end{equation}

\noindent The \textit{quantum-mechanical} expression for the radiative power%
\cite{8} associated with a transition between quantum states $nl$ and $
n^{\prime }l^{\prime }$ is like Eq. (10) except for the radial matrix
element,

\begin{equation}
\left\langle nl|r|n^{\prime }l^{\prime }\right\rangle =\int_{0}^{\infty }%
\frak{R}_{nl}(r)r\frak{R}_{n^{\prime }l^{\prime }}(r)dr,  \tag{12}
\end{equation}

\noindent in place of the transition amplitude $X_{nn^{\prime }}$. Here $%
\frak{R}_{nl}(r)$ is a radial wavefunction\cite{9} and $l$ denotes the
angular quantum number.

A pair of quantum numbers, $nl$, becomes necessary to characterize a quantum
state in Sommerfeld's extension of the Bohr model by elliptical orbits\cite
{10} as well as in quantum mechanics. An $\left( nl\right) $ Sommerfeld
ellipse has the same length of semimajor axis, the same binding energy, and
the same orbital frequency as the \textit{n}th Bohr orbit, Eqs. (1) - (3).
However, its semiminor axis is shorter,

\begin{equation}
b_{nl}=r_{{\small B}}n\sqrt{l(l+1)},  \tag{13}
\end{equation}

\noindent with $l=0,1,...,n-1$.

\section{RELATIONS WITH MATRIX ELEMENTS}

Figure 1 gives a comparison of quantum mechanically calculated matrix
elements, listed in Appendix A, with transition amplitudes between \textit{%
neighboring} Bohr orbits, $n\leftrightarrow n^{\prime }=n+1$. In this case,
after expansion and cancellation, Eq. (11) simplifies to

\begin{equation}
\frac{X_{nn^{\prime }}}{r_{{\scriptsize B}}}=n(n+1)+\frac{1}{3}. 
\tag{11$^{\prime }$}
\end{equation}

\noindent The value of the transition amplitude $X_{nn^{\prime }}$ is given
on the abscissa; the value of the matrix elements $\left\langle
nl|r|n^{\prime }l^{\prime }\right\rangle $ on the ordinate, together with a
repetition of the $X_{nn^{\prime }}$. This places the amplitudes $
X_{nn^{\prime }}$ of transitions between Bohr orbits, shown by circles, on
the diagonal line in the graph. The data align vertically in columns,
starting with the transition $n=1\leftrightarrow 2=n^{\prime }$ on the left,
and continue with transitions $2\leftrightarrow 3$, etc., until $
5\leftrightarrow 6$ on the right.

For a given pair of (principal) quantum-number neighbors, $n$ and $n^{\prime
}$, the matrix elements $\left\langle nl|r|n^{\prime }l^{\prime
}\right\rangle $ turn out to be always less than the corresponding
transition amplitudes $X_{nn^{\prime }}$. The transition amplitude, Eq.
(11), thus serves as an \textit{upper bound} for the respective matrix
elements.

The selection rule for dipole transitions, $\Delta l=\pm 1$, permits several
possible transitions\cite{11} between states with the same principal quantum
numbers, $n\leftrightarrow n^{\prime }$, depending on the states' angular
quantum numbers $l$ and $l^{\prime }=l\pm 1$. In Fig. 1 the matrix elements
of such transitions, $\left( n,l\right) \leftrightarrow \left( n+1,l\pm
1\right) ,$ fall beneath each transition amplitude $X_{nn^{\prime
}}=X_{n,n+1}$, forming the columns under the diagonal. Let us call the
transitions where \textit{both} quantum numbers increase or decrease, $%
\left( n,l\right) \leftrightarrow \left( n+1,l+1\right) $, \textit{comutant}%
\cite{12}. Their matrix elements are displayed by pointed area symbols ($%
\mathsf{\Box }$, $\Diamond $, $\triangle $) and connected with dashed trend
lines. In contradistinction, we want to call the transitions with \textit{%
oppositely} changing quantum numbers, $\left( n,l\right) \leftrightarrow
\left( n+1,l-1\right) $, \textit{contramutant}. Their matrix elements are
displayed by line symbols (\textsf{x}, +, $*$) and connected with dotted
trend lines. All comutant matrix elements end up above the dash-dotted line
in Fig. 1---the ``separatrix''---all contramutant matrix elements beneath.

The \textit{top} matrix element in each column---just beneath the
diagonal---represents the comutant transition between quantum states of
neighboring principal quantum numbers and the \textit{maximum} \textit{
angular} quantum numbers, $\left\langle nl|r|n+1,l+1\right\rangle
=\left\langle n,n-1|r|n+1,n\right\rangle $. In terms of Sommerfeld orbits
those quantum states are neighbor ellipses with the largest possible minor
axis, Eq. (13). Their semimajor axes are as long as the radii of the
respective Bohr orbits, Eq. (1), but their semiminor axes are slightly
shorter. In other words, they have the smallest possible deviation from
circularity that quantization permits. Figure 2(a) shows an example.

Proceeding down a given column in Fig. 1---from beneath the diagonal to
above the separatrix---we find the comutant matrix elements $\left\langle
nl|r|n+1,l+1\right\rangle $, with $l=n-1,...,0$. The Sommerfeld ellipses
involved in that descending order are progressively more slender, see Fig.
2(b), culminating in a line ellipse of the smaller orbit, $\left( n,0\right) 
$---(not shown in Fig. 2).

The \textit{bottom} matrix element in each column of Fig. 1 represents the 
\textit{contramutant} transition between quantum states of neighboring
principal quantum numbers and the \textit{maximum angular} quantum numbers, $%
\left\langle nl|r|n+1,l-1\right\rangle =\left\langle
n,n-1|r|n+1,n-2\right\rangle $. The corresponding Sommerfeld ellipses again
have the smallest possible deviation from circularity that quantization and
contramutant transition permit. But now the neighboring Sommerfeld orbits
are a short fat ellipse and a long slim ellipse confocally penetrating the
former one---see Fig. 2(c) for an example.

Proceeding upward a given column in Fig. 1, from the bottom to below the
separatrix, we find the contramutant matrix elements $\left\langle
nl|r|n+1,l-1\right\rangle $, with $l=n-1,...,1$. The Sommerfeld ellipses
involved in that ascending order are again progressively more slender; this
time culminating in a line ellipse of the larger orbit, $\left( n+1,0\right) 
$---(not shown in Fig. 2).

Why are the matrix elements $\left\langle nl|r|n^{\prime }l^{\prime
}\right\rangle $ always smaller than the corresponding transition amplitudes 
$X_{nn^{\prime }}$? What affects their value? And why are the matrix
elements for transitions between orbits with maximum circularity at both the
top and bottom of the columns in Fig. 1, but those with minimum circularity
next to the separatrix? The deviation of the matrix elements from the
transition amplitudes between Bohr orbits can conceptually be understood in
terms of orbit geometry and transition trajectory. We will find that orbit
geometry provides a scaling whereas the transition trajectory gives rise to
an interference effect.

Consider an electron orbiting along an $\left( nl\right) $ Sommerfeld
ellipse. As shown in Appendix B, its instantaneous acceleration consists of
a radial part and a ``centripetal'' part,

\begin{equation}
\left| \frac{d^{2}\mathbf{r}_{nl}}{dt^{2}}\right| =\frac{d^{2}r_{nl}}{dt^{2}}
-r_{nl}\left( \frac{d\alpha _{nl}}{dt}\right) ^{2}.  \tag{14}
\end{equation}

\noindent Furthermore, the electron's \textit{average} centripetal
acceleration is proportional to the orbit's semiminor axis,

\begin{equation}
\left\langle r_{nl}\left( \frac{d\alpha _{nl}}{dt}\right) ^{2}\right\rangle
=4\pi ^{2}f_{n}^{2}b_{nl}.  \tag{15}
\end{equation}

We first discuss the top matrix element in each column in Fig. 1, just below
the diagonal. It represents a comutant transition between neighbor ellipses
with the least deviation from circularity. For these orbits we will, in an
approximate treatment, consider only the average centripetal acceleration,
neglecting the radial contribution. By Eqs. (15), (13) and (1) the ratio of
acceleration in an $\left( nl\right) $ Sommerfeld ellipse and the $n$th Bohr
orbit is

\begin{equation}
\frac{\left\langle r_{nl}\left( \frac{d\alpha _{nl}}{dt}\right)
^{2}\right\rangle }{\left\langle r_{n}\left( \frac{d\alpha _{n}}{dt}\right)
^{2}\right\rangle }=\frac{b_{nl}}{r_{n}}=\frac{\sqrt{l(l+1)}}{n}.  \tag{16}
\end{equation}

According to Bohr's first postulate, no radiation is emitted or absorbed
while the electron keeps orbiting along the $\left( nl\right) $ or $\left(
n^{\prime }l^{\prime }\right) $ Sommerfeld ellipse. However, emission or
absorption occurs for an $\left( nl\right) \leftrightarrow \left( n^{\prime
}l^{\prime }\right) $ transition. It is reasonable to expect that the
average acceleration during the \textit{transition} is some average of the
average acceleration in both orbits. In the present approximation we employ
the simplest average---the arithmetic mean\cite{13}. It yields for a
comutant, $\left( nl\right) \leftrightarrow \left( n+1,l+1\right) $
transition

\begin{equation}
\frac{\left\langle n,l|r|n+1,l+1\right\rangle }{X_{n,n+1}}\approx \frac{1}{2}
\left\{ \frac{\sqrt{l(l+1)}}{n}+\frac{\sqrt{(l+1)(l+2)}}{(n+1)}\right\} . 
\tag{17a}
\end{equation}

\noindent For the top matrix element in each column of Fig. 1, Eq. (17a) is
95\% accurate or better. The approximation improves with increasing $n$---to
the right in Fig. 1---but gets worse with decreasing $l$---down toward the
separatrix. It ceases for transitions that involve orbits with $l=0$ (line
ellipses)---next to the separatrix.

In Fig. 1 the top matrix elements fall on a slightly concave trend-curve
which approaches the diagonal line of the corresponding transition
amplitudes $X_{n,n+1}$. In the large-$n$ limit where Bohr's correspondence
principle holds, the top matrix elements merge with the transition
amplitudes between Bohr orbits, $\left\langle n,n-1|r|n+1,n\right\rangle
\cong X_{n,n+1}$. This is also obtained from the approximation (17a) in the
large-$n$ limit where $l\cong l+1=n\cong n+1$.

A visualization of \textit{comutant} transitions between neighbor orbits is
facilitated by Fig. 2(ab). The electron's trajectory during a transition
between Bohr orbits (dashed) must be some spiral (not shown) between those
circles. Similarly, the transition trajectory between the Sommerfeld orbits
must be an elliptical spiral, connecting smoothly the outer and inner
ellipse. Note that the larger ellipse completely encompasses the smaller
ellipse, akin to the larger Bohr orbit's complete enclosure of the smaller
one. Therefore a comutant transition between Sommerfeld ellipses can be
considered as essentially a transition between Bohr orbits but geometrically
scaled by the ratio of minor axes, Eq. (17a).

The situation is quite different for a \textit{contramutant} transition,
illustrated in Fig. 2(c). What is a simple inward spiral between Bohr
orbits, $4\rightarrow 3$, now becomes an ``exotic'' transition from the long
slim to the short fat Sommerfeld orbit, $(4,1)\rightarrow (3,2)$, where the
electron has to move \textit{outside} the larger ellipse to reach the
smaller one. Thus, in contrast to transitions between Bohr orbits, where the
intra-orbital trajectory constructively contributes to the emission or
absorption of radiation, the extra-orbital trajectory in a contramutant
transition diminishes the radiation through partial cancellation. This
leads, qualitatively, to small values of the contramutant matrix elements,
falling beneath the separatrix in Fig. 1.

The pattern in Fig. 1, where the matrix elements at both the top \textit{and}
bottom of each column originate from transitions between the fattest
ellipses and those toward the separatrix from gradually slimmer ellipses,
suggests that the minor-axis scaling holds not only for comutant
transitions, Eq. (17a), but also for the contramutant transitions. Both
these influences---orbit scaling and cancellation due to extra-orbital
transition trajectory---are contained in an empirical approximation for the
contramutant matrix elements,

\begin{equation}
\frac{\left\langle n,l|r|n+1,l-1\right\rangle }{X_{n,n+1}}\approx \kappa
\left\{ \frac{\sqrt{l(l+1)}}{n}-\frac{\sqrt{(l-1)l}}{n+1}\right\} , 
\tag{17b}
\end{equation}

\noindent with a fudge factor $\kappa =1/4$. The formula is not derived from
any principles;\cite{13} it is devised in analogy to Eq. (17a) but with a
negative contribution for the long slim ellipse. It approximates the bottom
matrix elements reasonably well---except $(2,1)\leftrightarrow (3,0)$ which
involves a line ellipse. This finding may lend support to the notion of
counter-radiative effects from extra-orbital transition trajectories.

To demonstrate both approximations we compare the transition amplitude $
X_{34}\simeq 12.3$, Eq. (11$^{\prime }$), with the two largest matrix
elements in the third column of Fig. 1, that is, $\left\langle
32|r|43\right\rangle $ and $\left\langle 31|r|42\right\rangle $, and with
the bottom member, $\left\langle 32|r|41\right\rangle $. Their fraction of $
X_{34}$ is 83\%, 61\%, and 11\%, respectively. The scaled fractions, Eqs.
(17ab), are 84\%, 54\%, and 12\%. Figure 2(abc) shows the corresponding
elliptical orbits, together with the (dashed) Bohr orbits, for an assessment
of intra-orbital and extra-orbital transition trajectories.

While Eqs. (17ab) approximate well the top and bottom matrix elements in the
columns of Fig. 1, their accuracy deteriorates for the matrix elements
toward the separatrix. The reason is the increasing slenderness of the
involved ellipses, culminating in line orbits. The ellipses' slenderness
gives rise to stronger overtones (higher Fourier coefficients) of the radial
oscillations whose contribution to the acceleration, Eq. (14), have been
neglected in the minor-axis scaling, Eq. (16).

Going beyond the inspection of Sommerfeld ellipses, more quantitative
insight into matrix elements is obtained by the shape of the radial
wavefunctions, shown, for the three above cases, in Fig. 3(abc). The heavy
curve displays the integrand of Eq. (12); the sum of positive (negative)
areas between the radial axis and the curve above (beneath) visualizes the
matrix element. In a sense the matrix elements can be regarded as resulting
from interference of the weighted wavefunctions---constructive in the case $
32\leftrightarrow 43$, less so for $31\leftrightarrow 42$, and considerably
destructive for $32\leftrightarrow 41$. It may well be that such
wavefunction interference and the scaled trajectory effects, considered
above, are merely different manifestations of the same radiation dynamics.

\section{HISTORICAL PERSPECTIVE}

Max Born\cite{14} came close to the present approach of continuous changes
inside the atom with his observation that differential quotients in the
large-$n$ limit of quantum transitions correspond to difference quotients in
the small-$n$ regime,

\begin{equation}
\frac{\Delta \Gamma }{\Delta n}\Longleftrightarrow \frac{d\Gamma }{d\tilde{n}%
}.  \tag{18}
\end{equation}

\noindent The quantity $\Gamma $ is differentially related to a classical
(continuous) orbital quantity,

\begin{equation}
g(\tilde{n})\equiv k\frac{d\Gamma (\tilde{n})}{d\tilde{n}},  \tag{19}
\end{equation}

\noindent where $k$ is a coefficient of proportionality. The quantization of 
$g(\tilde{n})$ approaches in the large-$n$ limit the transition quantity

\begin{equation}
g_{nn^{\prime }}\approx g_{n}\Delta n=k\left. \frac{d\Gamma (\tilde{n})}{d%
\tilde{n}}\right| _{n}\Delta n,\quad n\gg \Delta n=n^{\prime }-n.  \tag{20}
\end{equation}

\noindent This is Bohr's correspondence principle---a generalization of Eq.
(6). With the analogy (18), called ``\textit{Born's} correspondence rule,''%
\cite{15} Eq. (20) generalizes to

\begin{equation}
g_{nn^{\prime }}=k\frac{\Delta \Gamma }{\Delta n}\Delta n=k\Delta \Gamma 
\tag{21}
\end{equation}

\noindent for any quantum number $n$. What Born didn't do was \textit{%
integrate} Eq. (19) to obtain the numerator of the difference quotient,

\begin{equation}
\Delta \Gamma =\int_{n}^{n^{\prime }}d\Gamma =\int_{n}^{n^{\prime }}\frac{%
d\Gamma (\tilde{n})}{d\tilde{n}}d\tilde{n}=k^{-1}\int_{n}^{n^{\prime }}g(%
\tilde{n})d\tilde{n},  \tag{22}
\end{equation}

\noindent and thus the transition property as an integral over the
corresponding orbital quantity,

\begin{equation}
g_{nn^{\prime }}=\int_{n}^{n^{\prime }}g(\tilde{n})d\tilde{n}.  \tag{23}
\end{equation}

\noindent Two specific examples of Eq. (23) are the above Eqs. (5') and (11).

\section{BIRTH OF A PHOTON}

It is tempting to determine the transition analogues of other orbital
quantities of the old Bohr model, such as the period of revolution,

\begin{equation}
T_{n}=\frac{1}{2}\frac{h}{R_{y}}n^{3},  \tag{24}
\end{equation}

\noindent and the orbital speed,

\begin{equation}
v_{n}=\frac{\alpha c}{n}.  \tag{25}
\end{equation}

\noindent Here $\alpha \approx 1/137$ is the fine-structure constant and $c$
is the speed of light. By Eq. (23) the corresponding transition period is

\begin{equation}
T_{nn^{\prime }}=\frac{1}{8}\frac{h}{R_{y}}(n^{\prime 4}-n^{4})  \tag{26}
\end{equation}

\noindent and the transition speed

\begin{equation}
v_{nn^{\prime }}=\alpha c\ln \left( \frac{n^{\prime }}{n}\right) .  \tag{27}
\end{equation}

\noindent What is the meaning of these quantities?

The transition period turns out to be slightly longer than the radiation
period, both being bracketed by the period of revolution of the involved
orbits, $T_{n}<1/f_{nn^{\prime }}<T_{nn^{\prime }}<T_{n^{\prime }}$. The
largest discrepancy holds for the $1\leftrightarrow 2$ transition, with a
ratio of $T_{nn^{\prime }}/f_{nn^{\prime }}^{-1}\simeq 1.4$ . In the limit
of transitions between high-quantum number orbits, $n$ and $n^{\prime }\gg
\Delta n$, all those periods merge, in accordance with the correspondence
principle.

Classical electrodynamics distinguishes between radiation phenomena near the
source of accelerating charges---the so-called ``near zone''---and those
very far from the source---the ``radiation zone.'' Near-zone effects are
instantaneously caused by changes of the source; far-zone effects are
retarded. Clearly, the transition frequency $f_{nn^{\prime }}$ and
transition amplitude $X_{nn^{\prime }}$, which together compose the
radiative power $\left\langle S_{nn^{\prime }}\right\rangle $, Eq. (10),
must be quantities of the radiation zone. In contrast, it seems likely that
the transition period $T_{nn^{\prime }}$ relates to the near zone between
orbits $n$ and $n^{\prime }$.

That conclusion can hardly be avoided for the transition speed $
v_{nn^{\prime }}$. For the $2\rightarrow 1$ inward electron transition Eq.
(27) gives rise to an outward transition speed,\cite{3} $|v_{21}|=\left|
\alpha c\ln \left( 1/2\right) \right| \simeq \left| -0.69\ v_{1}\right| ,$
that is, about 70\% of the electron's ground-state speed $v_{1}=\alpha c$,
Eq. (25). The transition speed $v_{nn^{\prime }}$ is very slow when the
electron transition occurs between high neighbor orbits but very fast when
the electron transits from a high orbit to the ground state. However, for
all practical purposes $v_{nn^{\prime }}$ will not exceed the speed of light 
$c.$\cite{16} These findings suggest that, with an inward transition of the
electron, the transition speed $v_{nn^{\prime }}$ is the (negative)\cite{3}
average radial\textit{\ speed of the nascent photon} in the near zone, that
is, between electron orbits $n$ and $n^{\prime }.$ By this interpretation
the photon starts from rest, $v=0$, at the beginning of the electron
transition. The transition period $T_{nn^{\prime }}$ can be regarded as the
time interval during which the nascent photon ``peels off'' (decouples) from
the inward spiraling electron. The fresh photon will keep accelerating
beyond the near zone until it reaches the speed of light $c$ in the
radiation zone.

When directions are reversed, the same scenario must describe the ``death''
of an absorbed photon. An incoming photon of radiation frequency $f_{12}$,
for instance, will decelerate as it approaches the near zone. The transition
speed, $v_{21}=\alpha c\ln (2/1)\simeq +0.69\ v_{1}$, represents the
(inward) average radial speed of the moribund photon between electron orbits
2 and 1.

\section{EPISTOMOLOGY}

Despite its initial successes (Balmer formula, space quantization,
fine-structure formula) the orbit-based old quantum theory of Bohr and
Sommerfeld had been insufficient in regard to the atom's magnetic properties
(Zeeman effect), the intensity of the spectral lines, the stability of the
hydrogen-molecule ion, $H_{2}^{+}$, and the $He$ atom. As we know now, with
the benefit of hindsight, one of the reasons for these shortcomings was the
ignorance of electron spin. However, in the early 1920s it was suspected,
chiefly by Pauli, Heisenberg and Born\cite{17}, that the failures of the old
quantum theory were caused by the fallacy of the very concept of electron
orbits.

In his article on matrix mechanics Heisenberg\cite{18} categorically
rejected the notion of electron orbits as unobservable in principle. In the
spirit of positivist philosophy he instead proposed that any theory in
physics should involve only relationships between fundamentally observable
quantities, such as frequency and intensity of spectral lines. The present
approach---a ``\textit{new} old quantum theory''---obtains both spectral
frequencies and intensities from electron orbits. How is that possible?

It has been pointed out\cite{19} that Heisenberg disobeyed his own demand by
invoking fundamentally unobservable quantities---virtual oscillators---in
his theory. Something similar occurs in Schr\"{o}dinger's wave mechanics
where wavefunctions $\Psi $ play a central role but are, by themselves,
unobservable. The orbit conundrum is readily resolved, though, if we regard
quantum orbits not as observable spatial descriptions---the notion of ``ring
atoms'' in the Bohr model or ``needle atoms'' for $l=0$ Sommerfeld orbits
contradicts all experience---but merely as entities to \textit{calculate}
observable quantities. This interpretation gives such \textit{virtual orbits}
in the new-old quantum theory a status equivalent to the virtual oscillators
in matrix mechanics or to the wavefunctions in wave mechanics. It also
renders the oft-mentioned incompatibility of quantum orbits with the
Heisenberg uncertainty principle immaterial.

The present modification of the Bohr model, with the first postulate in
place but the second postulate omitted, regards both the orbital and
transitional motion of the electron as \textit{continuous} processes. They
differ merely in the action variable,\cite{20} alluded to in the
introduction. Motion on an orbit trajectory is accompanied by constant
action, $I_{n}=nh/2\pi $, and contrary, motion on a transition trajectory by
continuously changing action, $I(\tilde{n})=\mathit{\tilde{n}}h/2\pi $. In
place of Bohr's two postulates we can rephrase their essence more
succinctly: \textit{Electrodynamic phenomena occur only in processes with
continuously changing action variable}. This automatically exempts the
stationary states from electrodynamics, restricting the latter to quantum
transitions. The use of a continuous quantum variable, $\tilde{n}$, rules
out quantum leaps and instead permits calculus and simple quantum
electrodynamics right in the heart of the atom.\bigskip

\noindent \textbf{ACKNOWLEDGMENTS}

\noindent I thank Ernst Mohler for valuable discussions. I also thank
Preston Jones and Van Katkanant for help with computer integration and
graphics.

\appendix 

\section{MATRIX ELEMENTS}

$
\begin{array}{lllll}
&  &  &  & \left\langle 54|27.21|65\right\rangle \\ 
&  &  & \left\langle 43|17.72|54\right\rangle & \left\langle
53|22.57|64\right\rangle \\ 
&  & \left\langle 32|10.23|43\right\rangle & \left\langle
42|14.06|53\right\rangle & \left\langle 52|18.58|63\right\rangle \\ 
& \left\langle 21|4.74|32\right\rangle & \left\langle 31|7.56|42\right\rangle
& \left\langle 41|11.0|52\right\rangle & \left\langle 51|14.2|62\right\rangle
\\ 
\left\langle 10|1.29|21\right\rangle & \left\langle 20|3.07|31\right\rangle
& \left\langle 30|5.47|41\right\rangle & \left\langle 40|8.5|51\right\rangle
& \left\langle 50|11.6|61\right\rangle
\end{array}
$

$
\begin{array}{lllll}
\quad -\cdot - & \quad \quad -\cdot - & \quad \quad -\cdot - & \quad \quad
-\cdot - & \quad \quad -\cdot - \\ 
\ \ \ \ \ \ \ \ \ \ \ \ \ \ \  & \left\langle 21|0.95|30\right\rangle & 
\left\langle 31|2.45|40\right\rangle \  & \left\langle
41|4.60|50\right\rangle \  & \left\langle 51|7.41|60\right\rangle \\ 
&  & \left\langle 32|1.30|41\right\rangle & \left\langle
42|3.02|51\right\rangle & \left\langle 52|5.43|61\right\rangle \\ 
&  &  & \left\langle 43|1.66|52\right\rangle & \left\langle
53|3.65|62\right\rangle \\ 
&  &  &  & \left\langle 54|2.02|63\right\rangle
\end{array}
$

TABLE I. Matrix elements $M$ for dipole transitions between quantum states $%
nl$ and $n^{\prime }l^{\prime }$, Eq. (12), here listed as $\left\langle
nl|M|n^{\prime }l^{\prime }\right\rangle $. Values, in the unit of Bohr
radius $r_{B}$, are from Ref. 21 or otherwise calculated by integration of
radial wavefunctions from Ref. 22. The dash-dotted line corresponds to the
separatrix in Fig. 1, with comutant transitions above and contramutant
transitions below.

\section{KEPLER ACCELERATION}

Consider Kepler motion of a body along an elliptical orbit. The body's
Cartesian coordinates are $x=r\cos \alpha $ and $y=r\sin \alpha $. The
components of its acceleration are

\begin{equation}
\frac{d^{2}x}{dt^{2}}=\left[ \frac{d^{2}r}{dt^{2}}-r\left( \frac{d\alpha }{dt%
}\right) ^{2}\right] \cos \alpha -[2\frac{dr}{dt}\frac{d\alpha }{dt}+r\frac{
d^{2}\alpha }{dt^{2}}]\sin \alpha  \tag{28a}
\end{equation}

\noindent and 
\begin{equation}
\frac{d^{2}y}{dt^{2}}=\left[ \frac{d^{2}r}{dt^{2}}-r\left( \frac{d\alpha }{dt%
}\right) ^{2}\right] \sin \alpha +[2\frac{dr}{dt}\frac{d\alpha }{dt}+r\frac{
d^{2}\alpha }{dt^{2}}]\cos \alpha  \tag{28b}
\end{equation}

\noindent By Kepler's second law the areal speed, here expressed in relation
to angular momentum $L$ and mass $m$ of the body, $2dA/dt=L/m\equiv C$, is a
constant of the motion,

\begin{equation}
C=r^{2}\frac{d\alpha }{dt}.  \tag{29}
\end{equation}

\noindent Its derivative, $dC/dt=0=r[2(dr/dt)(d\alpha /dt)+r(d^{2}r/dt^{2})]$
, makes the brackets on the far right of Eqs. (28 ab) vanish. We square and
add Eqs. (28 ab), then take the root,

\begin{equation}
\sqrt{\left( \frac{d^{2}x}{dt^{2}}\right) ^{2}+\left( \frac{d^{2}y}{dt^{2}}
\right) ^{2}}=\left| \frac{d^{2}\mathbf{r}}{dt^{2}}\right| =\frac{d^{2}r}{
dt^{2}}-r\left( \frac{d\alpha }{dt}\right) ^{2}.  \tag{30}
\end{equation}

\noindent This gives the acceleration as the sum\cite{23} of a radial and a
``centripetal'' term. We square Eq. (29) and solve for the instantaneous
centripetal acceleration,

\begin{equation}
r\left( \frac{d\alpha }{dt}\right) ^{2}=\frac{C^{2}}{r^{3}}.  \tag{31}
\end{equation}

\noindent Combined with the path-average\cite{24} of the inverse cube radial
distance, taken over a Kepler orbit,

\begin{equation}
\left\langle r^{-3}\right\rangle _{s}=\frac{1}{a^{2}b},  \tag{32}
\end{equation}

\noindent and the expression for the constant of motion in terms of orbital
semiaxes and frequency,\cite{25}

\begin{equation}
C=2\pi \,ab\,f,  \tag{33}
\end{equation}

\noindent the \textit{average} centripetal acceleration is

\begin{equation}
\left\langle r\left( \frac{d\alpha }{dt}\right) ^{2}\right\rangle =4\pi
^{2}f^{2}b.  \tag{34}
\end{equation}

\noindent \bigskip

\appendix

\noindent \textbf{FIGURE CAPTIONS}

Fig. 1. Comparison of dipole matrix elements $\left\langle nl|r|n^{\prime
}l^{\prime }\right\rangle $ with transition amplitudes $X_{nn^{\prime }}$
between neighboring Bohr orbits. The circles on the diagonal give $
X_{nn^{\prime }}$. The dash-dotted line (``separatrix'') divides matrix
elements of comutant (above) and contramutant (below) quantum transitions
(see text).\medskip

Fig. 2. Bohr orbits (dashed) and Sommerfeld ellipses involved in $
n\leftrightarrow n^{\prime }=3\leftrightarrow 4$ quantum transitions. The
corresponding transition amplitude and matrix elements are displayed in the
third column of Fig. 1 with $X_{34}$ on the diagonal, (a) directly beneath,
(b) next down, and (c) at the bottom.\medskip

Fig. 3. Radial wavefunctions (light curves) of the quantum states in Fig. 2
and matrix-element integrand, Eq. (12), (heavy curve).


\begin{thebibliography}{99}
\bibitem{1}  A planetary system of an orbiting electron around a massive
central nucleus is, by classical electrodynamics, unstable due to radiative
loss according to the (centripetal) acceleration of the electron charge. No
such instability of atoms is observed though. Bohr resolved this conflict
with his postulate of stationary states: Defying electrodynamics, electrons
can move in certain orbits without radiative energy loss. An electron moving
in such an orbit is said to be in a stationary state.
\bibitem{3}  Consistent with integration rules it is natural to associate
the suffix $n$ in the radiation frequency $f_{nn^{\prime }}$ and in other
transition quantities---Eqs. (5), (5$^{\prime }$), (11), (26) and
(27)---with the electron's initial orbit and $n^{\prime }$ with its target
orbit. On the other hand, it is convenient to regard emission as a positive
entity and absorption as negative. Similarly, mathematical convention
designates radial outward motion as positive and inward motion as negative.
Unfortunately, because of the opposite radial motion of electron and photon
during the transition, the sign of the electromagnetic transition
quantities, obtained from those equations, is opposite to convienience or
convention.
\bibitem{4}  Note that the Planck-Einstein relation, Eq. (4), is not used in
the derivation of Eq. (5$^{\prime }$).
\bibitem{7}  If the dipole oscillation, Eq. (7), is expressed in terms of
complex exponentials, $exp(\pm i2\pi ft)$, then the rhs of Eqs. (9) and (10)
must be multiplied by 4.
\bibitem{9}  $\frak{R}_{nl}(r)$ is the ``stretched'' radial wavefunction, in
contrast to the ``true'' radial wavefunction $R_{nl}(r)$ from $\psi
(r,\theta ,\varphi )=R_{nl}(r)\Theta _{lm}(\theta )\Phi _{m}(\varphi )$.
They are related by $\frak{R}_{nl}=rR_{nl}$ .
\bibitem{12}  The terms ``co-mutant'' and ``contra-mutant'' are chosen to
signify changes of quantum numbers $n$ and $l$ in the same or opposite
direction, respectively.
\bibitem{13}  The purpose of Eqs. (17ab) is conceptual and only intended to
indicate scaling by orbit geometry and constructive or destructive effects
from transition trajectories.
\bibitem{16}  Take, for example, an electron transition from a very high
orbit $N$ to the ground state, $n^{\prime }=1$. By Eq. (27) the transition
speed equals the speed of light $c$ when $N=exp(1/\alpha )$. The radius of
that orbit is $r_{N}\approx 10^{109}m$---much larger than the universe!
\end{thebibliography}
\end{document}